\documentclass[sigconf,natbib=true]{acmart}

\usepackage{xspace}
\usepackage{graphicx}
\usepackage{bm}
\usepackage{marvosym}
\usepackage[ruled,linesnumbered]{algorithm2e}
\usepackage{multirow}
\usepackage{makecell}
\usepackage{tcolorbox}
\usepackage{balance}

\newcommand{\paratitle}[1]{\vspace{1.5ex}\noindent\textbf{#1}}
\newcommand{\ie}{\emph{i.e.,}\xspace}

\newcommand{\eg}{\emph{e.g.,}\xspace}

\newcommand{\ignore}[1]{}

\newcommand{\method}{GRSU\xspace}

\copyrightyear{2025}
\acmYear{2025}
\setcopyright{acmlicensed}\acmConference[SIGIR '25]{Proceedings of the 48th International ACM SIGIR Conference on Research and Development in Information Retrieval}{July 13--18, 2025}{Padua, Italy}
\acmBooktitle{Proceedings of the 48th International ACM SIGIR Conference on Research and Development in Information Retrieval (SIGIR '25), July 13--18, 2025, Padua, Italy}
\acmDOI{10.1145/3726302.3730080}
\acmISBN{979-8-4007-1592-1/2025/07}

\begin{document}

% \title{Interacting with Simulated Users in Conversational Recommendation via Searching with Generative Reward Model}
\title{Search-Based Interaction For Conversation Recommendation via Generative Reward Model Based Simulated User}

\author{Xiaolei Wang}
\affiliation{%
    \institution{Gaoling School of Artificial Intelligence, Renmin University of China}
    \city{Beijing}
    \country{China}
}
\email{wxl1999@foxmail.com}

\author{Chunxuan Xia}
\affiliation{%
    \institution{Gaoling School of Artificial Intelligence, Renmin University of China}
    \city{Beijing}
    \country{China}
}
\email{xiachunxuan@ruc.edu.cn}

\author{Junyi Li}
\affiliation{%
    \institution{Department of Computer Science, National University of Singapore}
    \city{Beijing}
    \country{Singapore}
}
\email{junyi_cs@nus.edu.sg}

\author{Fanzhe Meng}
\affiliation{%
    \institution{School of Information and Software Engineering, University of Electronic Science and Technology of China}
    \city{Chengdu}
    \country{China}
}
\email{mengfanzhe16@gmail.com}

\author{Lei Huang}
\affiliation{%
    \institution{Meituan}
    \city{Beijing}
    \country{China}
}
\email{huanglei45@meituan.com}

\author{Jinpeng Wang}
\affiliation{%
    \institution{Meituan}
    \city{Beijing}
    \country{China}
}
\email{wangjinpeng04@meituan.com}

\author{Wayne Xin Zhao\textsuperscript{\Letter}}
\affiliation{
    \institution{Gaoling School of Artificial Intelligence, Renmin University of China}
    \city{Beijing}
    \country{China}
}
\email{batmanfly@gmail.com}
\thanks{\Letter\ Corresponding author.}

\author{Ji-Rong Wen}
\affiliation{%
    \institution{Gaoling School of Artificial Intelligence, Renmin University of China}
    \city{Beijing}
    \country{China}
}
\email{jrwen@ruc.edu.cn}

\renewcommand{\shortauthors}{Xiaolei Wang et al.}

\begin{abstract}
Conversational recommendation systems~(CRSs) use multi-turn interaction to capture user preferences and provide personalized recommendations.
A fundamental challenge in CRSs lies in effectively understanding user preferences from conversations.
Previous research primarily focuses on the issue of insufficient contextual information in conversations.
They address this by introducing external knowledge sources, such as knowledge graphs, large language models~(LLMs), and conversational recommendation corpora.
Based on this, they design specific alignment strategies (\eg prompt learning and instruction tuning) to integrate such knowledge for user preference understanding and item recommendation. 
However, user preferences can be multifaceted and complex, posing significant challenges for accurate recommendations even with access to abundant external knowledge. 
While interaction with users can clarify their true preferences, frequent user involvement may lead to a degraded user experience.

To address this problem, we propose a \textbf{G}enerative
\textbf{R}eward model based \textbf{S}imulated \textbf{U}ser, named \textbf{\method}, for \textit{automatic} interaction with CRSs.
The simulated user provides feedback to the items recommended by CRSs, enabling them to better capture intricate user preferences through multi-turn interaction.
Inspired by generative reward models, we design two types of feedback actions for the simulated user: \ie \textit{generative item scoring}, which offers coarse-grained feedback, and \textit{attribute-based item critiquing}, which provides fine-grained feedback.
To ensure seamless integration, these feedback actions are unified into an \textit{instruction}-based format, allowing the development of a unified simulated user via instruction tuning on synthesized data.
With this simulated user, \textit{automatic} multi-turn interaction with CRSs can be effectively conducted.
Furthermore, to strike a balance between effectiveness and efficiency, we draw inspiration from the paradigm of \textit{reward-guided search} in complex reasoning tasks and employ beam search for the interaction process.
On top of this, we propose an \textit{efficient} candidate ranking method to improve the recommendation results derived from interaction.
Extensive experiments on public datasets demonstrate the effectiveness, efficiency, and transferability of our approach.
\end{abstract}




\maketitle

\section{Introduction}

Conversational recommender systems~(CRSs)~\cite{jannach2021survey,gao2021advances} aim to provide personalized recommendations through multi-turn conversations.
Typically, a CRS consists of two components: a \textit{recommender} component that provides recommendations based on the user preference from the conversation context, and a \textit{conversation} component that generates responses based on the conversation context and item recommendations from the recommender component.

\begin{table}[t]
    \centering
    \caption{
        An illustrative case of the complexity of user preference.
        We use blue to mark the entities mentioned by the user, italicized red for successful recommendations, and italicized gray for unsuccessful ones.
        The user expresses preference for both ``horror'' and ``comedy'' movies.
        Initially, the CRS focuses solely on the "horror" aspect in its recommendations. With feedback from our simulated user, it considers both aspects and generates a successful recommendation.
    }
    \resizebox{\linewidth}{!}{%
        \begin{tabular}{p{0.2\linewidth}p{0.8\linewidth}}
            \toprule
            \textbf{USER:}     & Hello! \\
            \textbf{SYSTEM:}    & What kind of movies do you like? \\
            \textbf{USER:}     & When I was younger I really enjoyed \textcolor{blue}{{the A Nightmare on Elm Street (1984)}}. \\
            \textbf{SYSTEM:}    & Oh, you like {scary} movies? I recently watched {{Happy Death Day (2017)}}. \\
            \textbf{USER:}     & I do enjoy some of the newer \textcolor{blue}{horror} movies that I have seen as well. I also like \textcolor{blue}{comedy} movies. \\
            \midrule
            \multirow{2}{*}{\begin{tabular}{@{}l@{}}\textbf{SYSTEM:}\\\textbf{(human)}\end{tabular}}    & Have a try for \textcolor{red}{\textit{Happy Death Day 2U (2019)}}. You will like it! \\
            \midrule
            \multirow{2}{*}{\begin{tabular}{@{}l@{}}\textbf{SYSTEM:}\\\textbf{(CRS)}\end{tabular}}      & I would recommend \textcolor{gray}{\textit{A Quiet Place (2018)}}, a popular horror movie. \\
            \multirow{2}{*}{\begin{tabular}{@{}l@{}}\textbf{USER:}\\\textbf{(simulated)}\end{tabular}}    & Thanks for your advice! But I would like to see more \textcolor{blue}{comedy} elements, like \textcolor{blue}{Happy Death Day (2017)}. \\
            \multirow{2}{*}{\begin{tabular}{@{}l@{}}\textbf{SYSTEM:}\\\textbf{(CRS)}\end{tabular}}      & I know, you should watch \textcolor{red}{\textit{Happy Death Day 2U (2019)}}, a sequel to Happy Death Day (2017). \\
            \bottomrule
        \end{tabular}%
    }
    \label{tab:intro}
\end{table}

To develop an effective CRS, it is crucial to understand user preferences from the conversation context.
A primary challenge lies in the brevity of typical conversations, which often consist of only a few sentences and lack sufficient contextual information for accurate user preference understanding.
To tackle this issue, prior work introduces external knowledge sources, such as knowledge graphs~\cite{li2023trea,zhang2023variational,dao2024broadening}, large language models~(LLMs)~\cite{yang2024unleashing,he2023large}, and conversational recommendation corpora~\cite{xie2024neighborhood,dao2024broadening}.
Building on these resources, they design specialized alignment strategies, such as prompt learning~\cite{dao2024broadening} and instruction tuning~\cite{yang2024unleashing}, to integrate the additional knowledge for improved user preference understanding and item recommendation.
However, recent work~\cite{wang2023rethinking} reveals that user preferences are often multifaceted and complex, making accurate recommendations challenging even with enriched knowledge.
For instance, as illustrated in Table~\ref{tab:intro}, a user prefers both ``horror'' and ``comedy'' movies.
However, the CRS considers only ``horror'' in its initial attempt, resulting in a suboptimal recommendation.
To address this problem, NBCRS~\cite{xie2024neighborhood} proposes a neighborhood-based approach.
It retrieves conversations similar to the current one and utilizes item frequencies from these similar conversations to generate recommendations.
While NBCRS can handle the complexity of user preferences by leveraging collective patterns, its dependence on similar conversations limits its ability to generalize to novel user preferences.
At its core, this problem stems from the \textit{single-turn} nature of existing CRSs, which generate recommendations based solely on the conversation history.
Without ongoing interaction with users, CRSs cannot fully discern their true preference from the conversation history.
However, frequent user involvement may negatively impact their experience~\cite{jin2021key}.
Similarly, recent studies on the evaluation of CRSs also highlight the limitations of single-turn interactions~\cite{wang2023rethinking,zhu2024llm,huang2024concept}.
They propose interactive evaluation methods and develop LLM-based simulated users to reduce the need for the involvement of real users.
Nevertheless, these simulated users depend on ground-truth user preferences, which are unavailable at inference time.
How to develop simulated users in the absence of ground-truth preference labels remains a significant challenge.

To address the aforementioned challenges, we aim to develop a \textit{label-free} simulated user for \textit{automatic} interaction with CRSs to help them understand complex user preference and improve recommendation results.
Our approach draws inspiration from the great success of \textit{reward models}~\cite{cobbe2021training,lightmanlet} in complex reasoning tasks, which evaluate the correctness of candidate reasoning trajectories without ground-truth answers to provide feedback for the task model to revise its reasoning process.
In particular, the recently proposed \textit{generative reward models}~\cite{zhang24generative,mahan2024generative} further unify generation and reward modeling.
They represent the reward as the probability of a specific token~(\eg ``Yes'' or ``No'').
That is, the reward is modeled in a generative way, which demonstrates better generalization ability.
For our simulated user in conversational recommendation, which also needs to provide \textit{generative, label-free} feedback to CRSs to facilitate recommendation refinement, it is feasible to develop it based on generative reward models.

To this end, in our paper, we propose a \textbf{G}enerative \textbf{R}eward model based \textbf{S}imulated \textbf{U}ser, named \textbf{\method}, for automatic interaction with CRSs.
Our simulated user can provide feedback through two designed actions inspired by generative reward models: \textit{generative item scoring} as coarse-grained feedback and \textit{attribute-based item critiquing} as fine-grained feedback.
These two actions are unified into the format of \textit{instruction}, enabling the development of a unified simulated user via instruction tuning.
Based on this, \textit{automatic} multi-turn interaction can be conducted between various CRSs and our simulated user.
Furthermore, to achieve a balance between effectiveness and efficiency, we draw inspiration from the paradigm of \textit{reward-guided search}~\cite{wan2024alphazero,snell2024scaling} in complex reasoning tasks to leverage beam search in the interaction process.
On top of this, we propose an \textit{efficient} candidate ranking method to improve the recommendation results derived from the interaction process.

To validate the effectiveness of our approach, we conduct experiments on public datasets.
Experimental results show that with our simulated user, general LLM-based CRSs without fine-tuning outperform several trained CRS models.
Notably, with just one round of interaction, our approach achieves superior performance compared to the best baseline model on the \textsc{INSPIRED} dataset, underscoring its efficiency.
Additionally, the simulated user trained on high-resource datasets can effectively support CRSs in low-resource datasets, highlighting the 
transferability of our approach.

% We summarize our main contributions as follows:

% (1) To the best of our knowledge, it is the first time that a simulated user has been developed for automatic interaction with CRSs to better understand complex user preference and improve the recommendation.

% (2) We design two actions for the simulated user to provide feedback from different granularity.
% We further unify these two actions into the format of instruction and develop a unified simulater user by fine-tuning an LLM with synthesized instruction data.

% (3) We propose to use beam search in the interaction process to generate candidate items and efficiently rank them to derive the final recommendation result.

% (4) Extensive experiments on public datasets demonstrate the effectiveness, efficiency, and transferability of our approach.
\section{Related Work}

Our work is related to the following two research directions.

\paratitle{Conversational recommendation.}
Conversational recommender systems~(CRSs) aim to provide item recommendations through multi-turn interaction.
One line of work~\cite{lei2020estimation,lei2020interactive,li2021seamlessly} focuses on the optimization of interaction policy.
They simplify the interaction to pre-defined actions (\eg asking questions or making recommendations) and handcrafted templates.
Based on this, they optimize CRSs to give accurate recommendations within as few turns as possible.
Another line of work~\cite{wang2022towards,wang2023improving,zhao2023alleviating} focuses on the elicitation and understanding of user preference in more free-form natural language conversations.
Since conversations usually lack sufficient contextual information, existing work introduces knowledge from external resources, such as knowledge graphs~\cite{dao2024broadening}, large language models~(LLMs)~\cite{he2023large,yang2024unleashing}, and conversational recommendation corpora~\cite{dao2024broadening,xie2024neighborhood}.
% Based on this, they design specific alignment strategies~(\eg prompt learning~\cite{dao2024broadening} and instruction tuning~\cite{yang2024unleashing}) to incorporate the introduced knowledge for user preference understanding and item recommendation.
% However, the user preference can be multifaceted and complex, making accurate recommendations challenging even with enriched knowledge.
Our work follows the second category and proposes a simulated user, which can automatically interact with CRSs to help them discern the true user preference from a complex conversation.
% However, user preference can be complex, making it hard to give accurate recommendations even if sufficient knowledge is provided.
% In this paper, we aim to address this issue by developing a simulated user to provide feedback for CRSs in an automatic interaction process.

\paratitle{Generative reward models.}
Reward models~\cite{cobbe2021training,lightmanlet} have become an emerging topic for solving complex reasoning tasks.
For example, in the commonly used ``Best-of-N'' strategy~\cite{cobbe2021training}, a task model first generates several candidate solutions, then a reward model ranks these candidates and selects the best one as the final prediction.
Recently, generative reward models~\cite{mahan2024generative,zhang24generative} have been proposed, which unify generation and reward modeling by representing reward as the probability of a specific token.
Based on this, critiques can be introduced in generation for better reward modeling~\cite{mahan2024generative,zhang24generative}.
In this work, we take inspiration from generative reward models to design the actions of our simulated user.
\section{Approach}

\begin{figure*}
	\includegraphics[width=\textwidth]{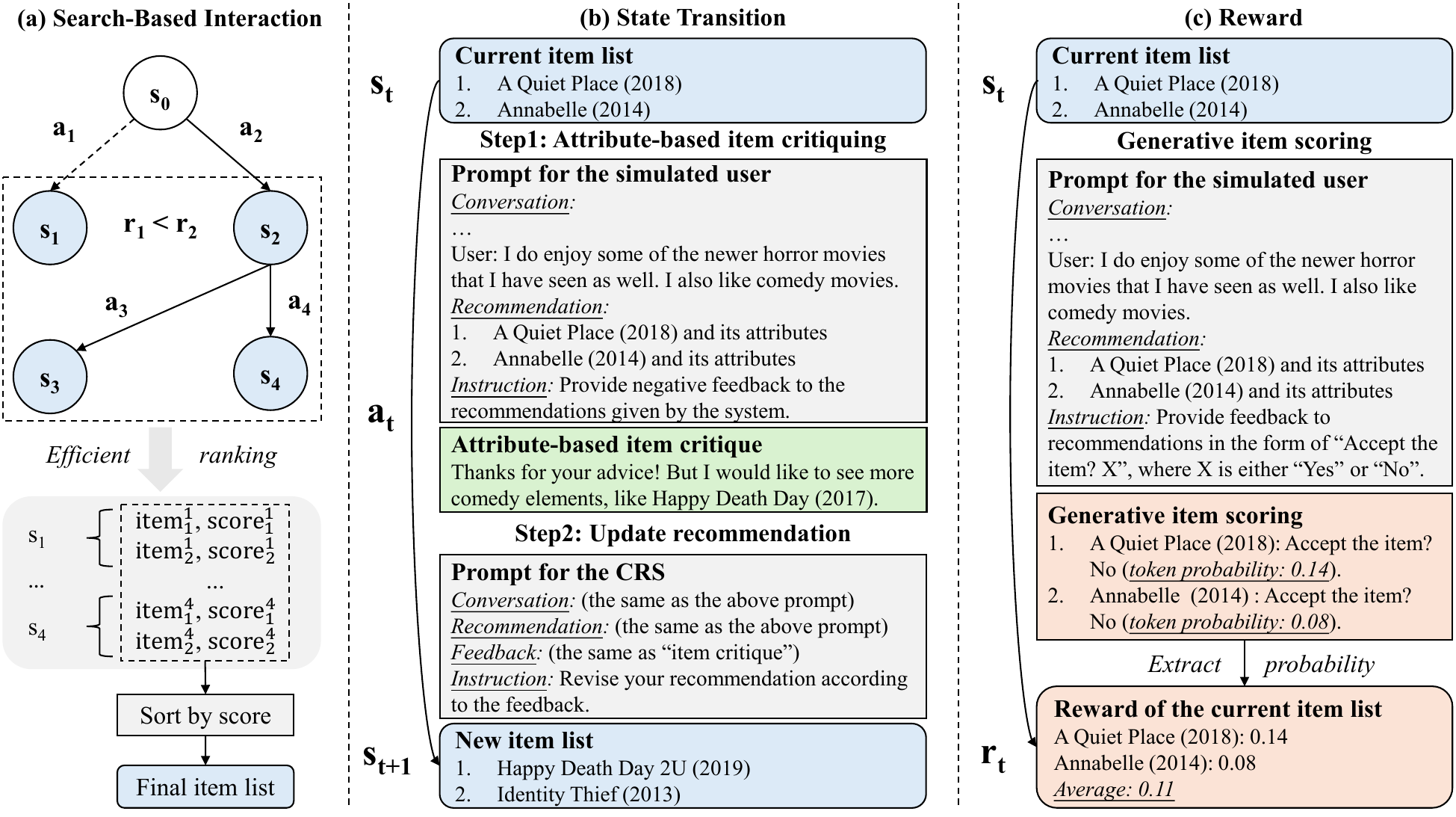}
	\centering
	\caption{
	    Approach overview.
	    (a) We use search-based interaction between CRSs and our simulated user to generate candidate items, which is followed by an efficient ranking method to derive the final item list.
		(b) A simplified state transition example: Starting from a current state (\ie an item list), the simulated user provides attribute-based item critiques, which the CRS uses to refine its previous recommendations, thereby transitioning to the next state.
		(c) A simplified reward example: Given a current state~(\ie an item list), the simulated user provides generative item scoring. The probability of the reward token (\ie ``No'') is extracted as the score for each item, and their average is the reward of the current state (\ie the item list).
    }
	\label{fig:method}
\end{figure*}

In this section, we present our simulated user based on generative reward models, name \textbf{\method}, for automatic interaction with CRSs, which is illustrated in Figure~\ref{fig:method}.

\subsection{Overview of the Approach}

\paratitle{Task formulation.}
Conversational recommender systems~(CRSs) aim to provide recommendations through multi-turn interaction.
At each turn, the system either makes recommendations or chats with the user.
Such a process ends when the user accepts recommended items or leaves.
A typical CRS comprises two core components: the recommender component and the conversation component.
In this paper, we focus on the \textit{recommender} component given the demonstrated prowess of LLMs in conversation~\cite{zhao2023survey,chang2024survey} following many LLM-based CRSs~\cite{he2023large,xie2024neighborhood}.
Formally, let $u$ demote a user and $i$ denote an item from the item set $\mathcal{I}$.
A conversation can be denoted as $C = \{ s_t, \mathcal{I}_t \}_{t=1}^\mathcal{T}$, where $s_t$ denotes the utterance at the $t$-th turn, and $\mathcal{I}_t$ denotes the items mentioned in $s_t$ ($\mathcal{I}_t$ is empty if no item is mentioned).
With the above definitions, the task of the recommender component in a CRS can be formalized as follows:
At the $k$-th turn, given the conversation history $C_{k-1} = \{ s_t, \mathcal{I}_t \}_{t=1}^{k-1}$, the recommender component needs to generate a ranked item list $\mathcal{\hat{I}}_k$ that best matches the ground-truth items in $\mathcal{I}_k$.

\paratitle{LLM-based CRSs.}
We use LLMs to build CRSs in our approach, as they have demonstrated strong performance in both item recommendation and response generation for conversational recommendation~\cite{he2023large,wang2023rethinking}.
Given that our idea is to improve CRSs through automatic interaction with simulated users, fine-tuning LLMs for this task is unnecessary.
Thus, we directly employ general-purpose LLMs as CRSs.
Specifically, following existing work~\cite{he2023large}, an LLM $M^r$ is prompted with a conversation history $C_{k-1}$, user feedback $f_{k}$ on the predicted recommendations $\mathcal{\hat{I}}_{k}$ (optional), and task description $d^r$, which can be represented as follows:
\begin{equation}
\label{eq:rec}
    \mathcal{\hat{I}}_k = \Phi^r(M^r(C_{k-1}, \mathcal{\hat{I}}_{k}, f_{k}, d^r)),
\end{equation}
where $\Phi^r$ is a post-processor that maps the items generated by the LLM $M^r$ to those in the item set $\mathcal{I}$.
Here, we use fuzzy matching\footnote{https://github.com/rapidfuzz/RapidFuzz} following existing work~\cite{he2023large}.
Notably, existing single-turn LLM-based CRSs~\cite{he2023large,yang2024unleashing,xie2024neighborhood} can be regarded as a special case of Eq.~\ref{eq:rec}, wherein only one recommendation is made, and the prompt only contains conversation history $C_{k-1}$ and task description $d^r$.

\paratitle{Interaction with simulated user.}
Our approach is inspired by the recent progress on \textit{LLM-based user simulation} for the evaluation of CRSs~\cite{wang2023rethinking,huang2024concept,zhu2024llm}, which utilizes LLMs for role-play with ground-truth user preferences.
In our approach, we aim to improve the recommendation of CRSs through automatic interaction with a \textit{label-free} simulated user.
As the basis, we fine-tune an LLM as the simulated user~(Section~\ref{sec:sim-user}).
It can provide feedback through two designed actions inspired by generative reward models: \ie \textit{generative item scoring} as coarse-grained feedback and \textit{attribute-based item critiquing} as fine-grained feedback.
With this simulated user, automatic interaction can be conducted with CRSs~(Section~\ref{sec:interaction}), where CRSs can iteratively refine their recommendations.
The approach of interaction with simulated users can be viewed as the paradigm of \textit{reward-guided search}~\cite{wan2024alphazero,snell2024scaling}.
Thus, we utilize beam search for interaction to balance effectiveness and efficiency.

\subsection{Generative Reward Model Based Simulated User}
\label{sec:sim-user}

The goal of simulated users in our approach is to provide feedback for CRSs, thereby enabling them to refine their recommendations.
Similarly, in complex reasoning tasks, the goal of reward models is to evaluate and provide feedback on reasoning trajectories, guiding task models toward more accurate solutions~\cite{cobbe2021training,lightmanlet}.
Since they share a similar target, it is feasible to develop our simulated user based on reward models.
In particular, we draw inspiration from \textit{generative reward models}~\cite{zhang24generative,mahan2024generative}, as the feedback provided to CRSs takes the form of natural language responses.
In this section, we first describe the two behaviors of our simulated user, followed by a detailed explanation of how to synthesize instruction data for each behavior to train the simulated user.

\subsubsection{User Behavior Description}
\label{sec:behavior}
To guide CRSs to effectively refine their recommendations, we draw inspiration from generative reward models to design two behaviors for the simulated user.

\paratitle{Generative item scoring.}
Item scoring~(\eg accepting or rejecting) is the most common behavior exhibited by users in conversational recommendation~\cite{cai2019towards,cai2020predicting}.
Existing work~\cite{yang2024unleashing,yang2022improving} imitates the item scoring behavior in a \textit{discriminative} way, \ie by learning to assign numerical scores to candidate items.
However, such an approach requires separate modeling of the item scoring behavior, as simulated users respond in a \textit{generative} way.
Similarly, in LLM-based reward models, most discriminative approaches~\cite{cobbe2021training,lightmanlet} add a separate module (\eg a linear head) to the LLM for reward modeling, which do not fully utilize the text-generation capabilities of LLMs.
To address this, a recent method, the \textit{generative reward model}~\cite{zhang24generative,mahan2024generative}, proposes representing the reward as the probability of generating a specific reward token (\eg ``\texttt{Yes}'' or ``\texttt{No}''), and adopting next-token prediction as the training objective.
This approach unifies reward and language modeling, which effectively leverages the text-generation capabilities of LLMs for reward modeling.
Inspired by this, we propose to model item scoring behavior in a \textit{generative} way.
Specifically, we frame item scoring as a single-choice question that requires the simulated user to select one option.
The predicted probability of the selected option token is then interpreted as the item score.
In addition, considering that CRSs give recommendations in the format of a list, we propose to score each item individually and take the average as the overall score for the item list.
This is easier to learn compared with directly scoring the entire list.
Formally, at the $k$-th turn, a generative item scoring $f^s_k$ from our simulated user $M^u$ can be represented as follows:
\begin{equation}
\label{eq:scoring}
    f^s_k = \frac{1}{| \mathcal{\hat{I}}_k^a |} \sum_{j=1}^{| \mathcal{\hat{I}}_k^a |} \Phi^s(M^u(C_{k-1}, {\hat{i}}_j^a, d^s)),
\end{equation}
where $C_{k-1}$ is the conversation history, $\mathcal{\hat{I}}_k^a$ is the list of items recommended by CRSs with their attributes, ${\hat{i}}_j^a$ is one of the items in the list with its attributes, $d^s$ is the task description, and $\Phi^s$ is a post-processor that extracts the probability of the reward token.
The task description employed for this behavior is as follows:
\textit{``You are the user to look for recommendations. You need to provide feedback on the recommendations given by the system. For each recommendation, you should give feedback in the form of ``Accept the recommendation (Yes/No)? X'', where X is either ``Yes'' or ``No''.''}

\paratitle{Attribute-based item critiquing.}
Critiquing is a representative method for eliciting user preferences using pre-defined attributes in early CRSs~\cite{chen2012critiquing}.
While these critiquing-based methods are straightforward to design, they inherently constrain the expressiveness of the user preference space.
Recently, LLMs have shown promise in simulating general human behaviors due to their generative nature~\cite{park2023generative}.
However, when employed as specific user simulators for conversational recommendation, LLMs exhibit limitations in aligning their preferences with those of real users~\cite{yoon2024evaluating}.
This problem also exists in reward models, which tend to predict rewards based on superficial features instead of real human preferences.
To solve this, natural language critiques have been introduced to calibrate the rewarding process~\cite{mahan2024generative,zhang24generative}.
Inspired by this, we propose to improve LLM-based user simulators with the \textit{critiquing} behavior.
Specifically, LLMs are prompted to provide critiques about the items recommended by CRSs.
Since the goal of simulated users is to assist CRSs in refining their recommendations, the critique should be specific and constructive.
To facilitate this, we add the attributes of the recommended items, enabling the simulated user to generate critiques from the perspective of \textit{attributes}.
Formally, at the $k$-th turn, an attribute-based item critique $f^c_k$ from our simulated user $M^u$ can be represented as follows:
\begin{equation}
\label{eq:critique}
    f^c_k = M^u(C_{k-1}, \mathcal{\hat{I}}_k^a, d^c),
\end{equation}
where $C_{k-1}$ is the conversation history, $\mathcal{\hat{I}}_k^a$ is the list of recommended items given by CRSs with their attributes, and $d^c$ is the task description (\textit{``You are the user to look for recommendations. Read the conversation and provide feedback on the recommendations given by the system.''}).
% The task description we used for this behavior is as follows:
% \textit{``You are the user to look for recommendations. Read the conversation and provide feedback to the recommendations given by the system.''}

\subsubsection{Instruction Data Synthesis For Model Training}

As discussed earlier, both user behaviors share the same contextual information (\ie conversation history and items recommended by CRSs) and differ only in the task description, as shown in Eq.~\ref{eq:critique} and~\ref{eq:scoring}.
This shared context allows for the development of a unified simulated user through instruction tuning, which involves fine-tuning an LLM using instruction data with distinct instructions for each user behavior.
Such an instruction-tuned LLM can effectively and efficiently facilitate the interaction with CRSs.
In this part, we first introduce how to synthesize instruction data and then describe how to utilize these data for model training.

\paratitle{Instruction Data Synthesis.}
Instruction data is usually composed of an instruction that contains the task description and its contextual information and a targeted output~\cite{zhang2023instruction}.
Here, we detail how to synthesize instruction data for the two user behaviors introduced in Section~\ref{sec:behavior}.

\textbullet~{\textit{Attribute-based item critiquing}}:
For this behavior, although real user feedback on recommendations is available in the dataset, it is often in the style of chit-chat and limited in information density (\eg lacking detailed item attributes)~\cite{wang2023rethinking}.
This limitation makes it challenging for CRSs to effectively refine their recommendations.
Thus, we opt to synthesize such instruction data using an LLM.
Specifically, for each recommendation turn $k$ in a conversation of the dataset, we take the conversation history $C_{k-1}$, recommended item list $\mathcal{\hat{I}}_k^a$, user preference $p_k$, and task description $d^c$ to construct the instruction $I^c_k = [C_{k-1}; \mathcal{\hat{I}}_k^a; p_k; d^c]$.
For the recommended item list $\mathcal{\hat{I}}_k^a$, to cover as many cases as possible, we adopt two kinds of construction methods.
Assuming the list contains \(l\) items, the first method incorporates ground-truth items by sampling \(l_p\) items (\(l_p \leq l\)) from the ground-truth set \(\mathcal{I}_k\) and \(l_n = l - l_p\) items as negatives from the remaining set \(\mathcal{I} - \mathcal{I}_k\). 
In contrast, the second method assumes no ground-truth items in the list, sampling \(l\) items exclusively from \(\mathcal{I} - \mathcal{I}_k\).
Compared to Eq.~\ref{eq:critique}, we incorporate a new element, \textit{user preference}, into the instruction, which includes ground-truth items \(\mathcal{I}_k\) and their attributes.
This addition aims to align the response more closely with the real user behavior.
Using the instruction \(I^c_k\), we prompt the LLM to generate the output \(o^c_k\).  
At inference time, however, ground-truth user preferences are unavailable. 
Thus, for instruction tuning, we modify the instruction to exclude this element, yielding \(I^{'c}_{\phantom{'}k} = [C_{k-1}; \mathcal{\hat{I}}_k^a; d^c]\). 
The resulting instruction is paired with \(o^c_k\) to create the training data.

\textbullet~{\textit{Generative item scoring}}:
For this behavior, we adopt the same instruction, $I^s_k = [C_{k-1}; {\hat{i}}_j^a; d^s]$, as defined in Eq.~\ref{eq:scoring} for data collection.
For the recommended item ${\hat{i}}_j^a$, corresponding ground-truth items $\mathcal{I}_k$ are already present in the dataset. Additionally, we sample $k$ negative items from the set $\mathcal{I} - \mathcal{I}_k$ for each ground-truth item as negative samples.
As for the output associated with $I^s_k$, we follow existing work of generative reward models~\cite{zhang24generative,mahan2024generative} to use a single-choice question $q$~(\ie ``\texttt{Accept the recommendation (Yes/No)?}'') concatenated with the option token $w$ (\ie ``\texttt{Yes}'' or ``\texttt{No}'').
This output is represented as $o^s_k = [q; w]$.

\paratitle{Model Training.}
With the constructed instruction data $\mathcal{D}$, \ie $\{( I^{'c}_{\phantom{'}k}, o^c_k ) \}_{k=1}^{n^c}$ and $\{(I^s_k, o^s_k)\}_{k=1}^{n^s}$, we can fine-tune an LLM as our simulated user through instruction tuning.
Since both instructions and target outputs are represented in natural language, the training objective can be unified under a sequence-to-sequence loss function.
Formally, we minimize the following loss function for model training:
\begin{equation}
    L = - \sum_{(I, o) \in \mathcal{D}} \sum_{i=1}^{|o|} \log \text{Pr}(o_i | o_{<i}, I ; \Theta^u),
\end{equation}
where $I$ and $o$ denote the instruction and output, respectively, $\Theta^u$ is the parameters of the LLM used as the simulated user, and $o_{<i}$ refers to the tokens preceding the $i$-th position.

\subsection{Search-Based Interaction Between CRS and Simulated User}
\label{sec:interaction}

With the simulated user introduced in the previous section, we can conduct multi-turn interaction with various CRSs.
Similarly, in complex reasoning tasks, reward models can be involved in the search process to guide task models toward the correct solution.
Drawing inspiration from this, we conceptualize the multi-turn interaction between CRSs and our designed simulated user as a \textit{search} process guided by the simulated user.
In this section, we first present the formalization of this interaction, followed by a detailed explanation of how to search for recommendation candidates and efficiently rank them to get the final recommendation result.

\paratitle{The formulation of interaction.}
Recall that LLM-based CRSs revise recommendations solely based on the most recent set of item recommendations and the feedback on them, as defined in Eq.~\ref{eq:rec}.
This makes it possible to model the interaction process as a Markov Decision Process~(MDP) with the tuple $(\mathcal{S}, \mathcal{A}, T, r)$.
We define the state $s_k \in \mathcal{S}$ as the item list $\mathcal{\hat{I}}_k$ recommended by CRSs and the action $a_k \in \mathcal{A}$ as the critique $f_k^c$ given by the simulated user (defined in Eq.~\ref{eq:critique}).
The transition function $T$ integrates the recommended items $\mathcal{\hat{I}}_k$ from the current state $s_k$ and the critique $f_k^c$ from the action $a_k$ into the recommendation prompt~(defined in Eq.~\ref{eq:rec}) for LLM-based CRSs to generate a new item list $\mathcal{\hat{I}}_{k+1}$ as the next state $s_{k+1}$.
The reward function $r_k = r(s_k, a_k)$ measures the quality of the action $a_k$ applied to the state $s_k$.
Since ground-truth labels are unavailable during inference, we use the score $f_k^s$ generated by the simulated user~(defined in Eq.~\ref{eq:scoring}) as the reward.

\paratitle{Searching the recommendation candidates.}
The formulation of multi-turn interaction as an MDP allows us to leverage advanced search algorithms.
Here, to balance simplicity and performance, we adopt beam search~\cite{wan2024alphazero} to optimize the recommendation candidates.
Specifically, we define a fixed beam width $B$, expansion width $N (N >= B)$, and search depth $D$.
In the beginning, we generate the initial states by sampling $N^2$ item lists following Eq.~\ref{eq:rec}, where we only include the conversation history and task description in the prompt.
During each search iteration, we proceed as follows: (1) the newly expanded states are scored using the reward function $r$; (2) only the top $B$ states with the highest rewards are retained; (3) each retained state is expanded into $N$ new states by sampling $N$ actions per state and applying the transition function $T$ to each $(\text{state}, \text{action})$ pair.
This process is repeated for $D$ iterations.

\paratitle{Efficient candidate ranking.}
The aforementioned search process produces multiple candidate item lists for recommendation.
However, these item lists may not necessarily be optimal.
Recall that the score of an item list is the average of its individual items, as stated in Section~\ref{sec:behavior}.
Consequently, a high reward for an item list does not guarantee that every item within the list is of high quality. 
Similarly, the item lists with lower rewards may still contain valuable items.
In light of this, to derive the most suitable item list from the search process, we propose to rank the items from all the generated item lists and take the top-$l$ items with the highest scores as the final item list for recommendation. 
Note that such a ranking method is \textit{efficient} because the scores for individual items are already computed during the search process, and the only additional step required is a sorting operation based on these scores.

\section{Experiment}

In this section, we first set up the experiments and then report the results and give a detailed analysis.

\subsection{Experimental Setup}

\begin{table}[t]
    \centering
    \caption{Statistics of the datasets after preprocessing.}
    \begin{tabular}{c|rrrr}
        \toprule
        \textbf{Datasets} & \textbf{\#Conversations} & \textbf{\#Turns} & \textbf{\#Users} & \textbf{\#Items} \\
        \midrule
        INSPIRED         & 999                      & 35,686           & 999            & 1,967            \\
        ReDial           & 11,348                   & 139,557          & 764            & 6,281           \\
        \bottomrule
    \end{tabular}
    \label{tab:dataset}
\end{table}

\paratitle{Datasets.}
To verify the effectiveness of our approach, we conduct experiments on two widely used conversational recommendation datasets, following existing work~\cite{he2023large,dao2024broadening,yang2024unleashing}, \ie \textsc{ReDial}~\cite{li2018towards} and \textsc{INSPIRED}~\cite{hayati2020inspired}.
The \textsc{ReDial} dataset is a conversational recommendation dataset about movie recommendations and is constructed by employing crowdsourcing workers on Amazon Mechanical Turk~(AMT).
Similar to \textsc{ReDial}, the \textsc{INSPIRED} dataset also centers on conversational movie recommendations, but with a much smaller size.
The statistics of both datasets are summarized in \tablename~\ref{tab:dataset}.
We use the same dataset splits as a recent LLM-based CRS work~\cite{he2023large,xie2024neighborhood}, which removes repeated items within each conversation for more reliable evaluation.

\paratitle{Baselines.}
We make comparisons with two groups of baselines:

(1) Conventional recommendation models:

\textbullet~{\textbf{Item Populalarity}}: It ranks items according to their frequencies of recommendation in the training set.

\textbullet~{\textbf{TextCNN}}~\cite{DBLP:conf/emnlp/Kim14}: It adopts a CNN-based model to extract textual features from conversation histories as the user embedding.

(1) General LLMs as zero-shot CRSs~\cite{he2023large}:

\textbullet~{\textbf{Llama-3.1-8B-Instruct}}~\cite{dubey2024llama}: It is a representative 8B parameter open-source model.

\textbullet~{\textbf{Phi-4}}~\cite{abdin2024phi}: It is a 14B parameter open-source model trained using strategically synthetic data and novel post-training methods, which surpasses GPT-4o on representative reasoning benchmarks.

\textbullet~{\textbf{Llama-3.3-70B-Instruct}}~\cite{dubey2024llama}: It is a representative 70B parameter open-source model.

\textbullet~{\textbf{GPT-4o}~\cite{hurst2024gpt}}: It is a versatile, high-intelligence flagship closed-source model.
We use the \texttt{gpt-4o-2024-08-06} version for it.

(2) Representative CRS models:

% \textbullet~{\textbf{KBRD}}~\cite{chen2019towards}: It utilizes an external knowledge graph to enhance the semantics of entities mentioned in the conversation.

% \textbullet~{\textbf{KGSF}}~\cite{zhou2020improving}: It incorporates two knowledge graphs to enhance the semantic representations of both words and entities mentioned in the conversation, and utilizes the Mutual Information Maximization method to align the semantic spaces of the two knowledge graphs.

% \textbullet~{\textbf{UniCRS}}~\cite{}: It designs knowledge-enhanced prompts based
% on the pre-trained DialoGPT model to fulfill the conversation and recommendation tasks in a unified approach.

\textbullet~{\textbf{TREA}}~\cite{li2023trea}: It constructs a multi-hierarchical, scalable tree for reasoning over the mentioned entities in the conversation.

\textbullet~{\textbf{VRICR}}~\cite{zhang2023variational}: It proposes a variational Bayesian method to dynamically refine the incomplete knowledge graph and select relevant knowledge.

\textbullet~{\textbf{DCRS}}~\cite{dao2024broadening}: It designs a demonstration retriever with knowledge-aware contrastive learning and an adaptive prompt learning approach to utilizing retrieved exemplars.

\textbullet~{\textbf{NBCRS}}~\cite{xie2024neighborhood}: It is a neighbourhood-based CRS that utilizes the frequency of items in similar conversations.

\textbullet~{\textbf{ReFICR}}~\cite{yang2024unleashing}: It fine-tunes an LLM as CRS with retrieval and generation instructions constructed from existing datasets.

\paratitle{Evaluation metrics.}
Given the demonstrated prowess of LLMs in conversation~\cite{chang2024survey}, we follow existing work~\cite{he2023large,xie2024neighborhood} to focus our evaluation on the performance of the recommendation subtask.
We adopt Recall@$k$, NDCG@$k$, and MRR@$k$ ($k$=10, 50) for performance evaluation.
For all the above metrics, we calculate and report the average scores on all the test examples from three runs.

\paratitle{Implementation details.}
For both CRSs and our simulated user, we set the length of all the item lists to 10 and utilize the attributes \textit{genre}, \textit{actor}, \textit{writer}, and \textit{director} for each item (movie).
In our main experiments, we employ \texttt{Phi-4} for LLM-based CRSs, as it offers a balance between excellent performance and appropriate model size. 
For detailed analysis, we additionally consider models with varying parameter scales, including \texttt{Llama-3.1-8B-Instruct}, \texttt{Llama-3.3-70B-Instruct}, and \texttt{GPT-4o}.
For the simulated user, we adopt \texttt{Llama-3.1-8B-Instruct} as the base model and leverage \texttt{Llama-3.3-70B-Instruct} for generating instruction data. 
During instruction data synthesis, we construct item lists for attribute-based item critiquing by including all possible subsets of ground-truth items with lengths less than 10 for lists containing ground-truth items. For item lists without ground-truth items, we sample one list per conversation history. For generative item scoring, we sample one additional item as a negative sample for each ground-truth item.
For instruction tuning, we train for one epoch with a batch size of 4. We utilize the AdamW optimizer with default parameters, a linear warmup over 1,000 steps, and a cosine decay learning rate scheduler, which decays to 10\% of the peak learning rate after the decay period. The learning rate is set to $10^{-6}$, with a weight decay of 0.01 and gradient norm clipping at 1.0.
For tuning LLMs, we use bf16 16-bit mixed precision training, FlashAttention-2~\cite{daoflashattention}, gradient checkpointing~\cite{chen2016training}, and DeepSpeed with ZeRO-3~\cite{rasley2020deepspeed}.
For searching recommendation candidates, we set the beam width and expand width to 4 and the search depth to 5. 
For the implementation of baseline methods, we use CRSLab~\cite{zhou2021crslab}.
Our code is available at the link: \url{https://github.com/RUCAIBox/GRSU}.

\subsection{Main Results}

% \usepackage{graphicx}
\begin{table*}
    [t]
    \centering
    \caption{ Main results of our approach and baseline methods. Following
    existing work~\cite{he2023large,xie2024neighborhood}, we remove repeated items within each conversation
    for more reliable evaluation. The best method is marked in bold. Numbers marked
    with * indicate that the improvement is statistically significant compared with
    the baseline (t-test with p-value < 0.05). }
    \label{tab:main_exp}
    % \resizebox{\textwidth}{!}{%
    \begin{tabular}{l|cccccc|cccccc}
        \toprule 
        \multicolumn{1}{c|}{\multirow{4}{*}{\textbf{Methods}}} & \multicolumn{6}{c|}{\textbf{ReDial}} & \multicolumn{6}{c}{\textbf{INSPIRED}} \\
        \cmidrule{2-13}
        & \multicolumn{2}{c}{\textbf{Recall}}  & \multicolumn{2}{c}{\textbf{NDCG}}    & \multicolumn{2}{c|}{\textbf{MRR}} & \multicolumn{2}{c}{\textbf{Recall}} & \multicolumn{2}{c}{\textbf{NDCG}} & \multicolumn{2}{c}{\textbf{MRR}} \\
        \cmidrule{2-13}                                                 
        & \textbf{@10}                         & \textbf{@50}                         & \textbf{@10}                      & \textbf{@50}                        & \textbf{@10}                      & \textbf{@50}                    & \textbf{@10} & \textbf{@50} & \textbf{@10} & \textbf{@50} & \textbf{@10} & \textbf{@50} \\
        \midrule 
        Populalarity           & 0.053 & 0.183 & 0.029 & 0.057 & 0.021 & 0.027 & 0.155 & 0.322 & 0.085 & 0.122 & 0.064 & 0.071 \\
        TextCNN                & 0.066 & 0.187 & 0.033 & 0.059 & 0.023 & 0.028 & 0.119 & 0.245 & 0.066 & 0.094 & 0.050 & 0.056 \\
        \midrule
        Llama-3.1-8B-Instruct  & 0.147 & 0.265 & 0.077 & 0.103 & 0.055 & 0.061 & 0.166 & 0.285 & 0.090 & 0.117 & 0.066 & 0.072 \\
        Phi-4                  & 0.173 & 0.313 & 0.095 & 0.126 & 0.071 & 0.077 & 0.145 & 0.277 & 0.077 & 0.106 & 0.056 & 0.062 \\
        Llama-3.3-70B-Instruct & 0.174 & 0.317 & 0.096 & 0.128 & 0.073 & 0.079 & 0.166 & 0.311 & 0.093 & 0.127 & 0.070 & 0.079 \\
        GPT-4o                 & 0.187 & 0.355 & 0.104 & 0.141 & 0.079 & 0.087 & 0.217 & 0.378 & 0.124 & 0.159 & 0.096 & 0.103 \\
        \midrule
        NBCRS                                                 & 0.130            & 0.287            & 0.065           & 0.104           & 0.044           & 0.052          & 0.123            & 0.204            & 0.068           & 0.085           & 0.057           & 0.057          \\
        VRICR                                                 & 0.148            & 0.327            & 0.066           & 0.105           & 0.041           & 0.049          & 0.082            & 0.244            & 0.039           & 0.075           & 0.026           & 0.034          \\
        TREA                                                  & 0.175            & 0.365            & 0.083           & 0.124           & 0.055           & 0.063          & 0.136            & 0.307            & 0.080           & 0.115           & 0.063           & 0.069          \\
        DCRS                   & 0.179 & 0.366 & 0.091 & 0.132 & 0.064 & 0.073 & 0.192 & 0.362 & 0.123 & 0.161 & 0.101 & 0.109 \\
        ReFICR                 & 0.216 & 0.467 & 0.099 & 0.155 & 0.065 & 0.076 & 0.147 & 0.326 & 0.074 & 0.113 & 0.051 & 0.060 \\
        \midrule
        \method                  & \textbf{0.250}*   & \textbf{0.488}*   & \textbf{0.135}*  & \textbf{0.187}*  & \textbf{0.100}*  & \textbf{0.111}* & \textbf{0.352}*   & \textbf{0.559}*   & \textbf{0.212}*  & \textbf{0.258}*  & \textbf{0.170}*  & \textbf{0.180}* \\
        \bottomrule
    \end{tabular}%
    % }
\end{table*}

Table~\ref{tab:main_exp} presents the recommendation performance of baseline and our methods.
% For the two conventional recommendation methods, Popularity demonstrates comparable performance to TextCNN on the \textsc{ReDial} dataset but outperforms TextCNN on the \textsc{INSPIRED} dataset.
% Popularity utilizes global item statistics for recommendation, while TextCNN utilizes textual features of conversations for recommendation.
% Since conversations usually have limited information density, the performance of TextCNN is heavily influenced by the dataset size.
% Thus, it achieves superior performance on the high-resource \textsc{ReDial} dataset compared to the low-resource \textsc{INSPIRED} dataset.
% Conversely, the global item statistics can be more informative than a single conversation for recommendation, which leads to better performance of Popularity on the \textsc{INSPIRED} dataset.
For general LLM-based zero-shot CRSs, the performance order is consistent across both datasets, \ie \texttt{GPT-4o} > \texttt{Llama-3.3-70B\-Instruct} > \texttt{Phi-4} > \texttt{Llama-3.1-8B-Instruct}.
This ranking aligns with their performance on other benchmarks.
Notably, while \texttt{Phi-4} performs closely to \texttt{Llama-3.3-70B-Instruct}, especially on the \textsc{ReDial} dataset, its model size is significantly smaller. 
This efficiency is why we employ \texttt{Phi-4}.

For representative CRS models, we can see that DCRS and ReFICR mostly outperform the other baseline methods in all the metrics, while the former performs better on the \texttt{INSPIRED} dataset and the latter performs better on the \texttt{ReDial} dataset.
Both methods leverage external knowledge to enhance the understanding of user preference from the conversation context.
Specifically, DCRS incorporates the knowledge from a conversational recommendation corpus by retrieval, while ReFICR utilizes the knowledge from an LLM by instruction tuning.
Given the abundant size of the \textsc{ReDial} dataset, ReFICR effectively utilizes the semantic and item knowledge encoded in the LLM for improved user preference understanding and item recommendation. 
This advantage enables it to outperform DCRS on this high-resource dataset.
However, on the low-resource \texttt{INSPIRED} dataset, the instruction tuning of ReFICR may be insufficient.
In contrast, retrieval from an existing corpus compensates for the data scarcity, which makes DCRS outperform ReFICR.

Lastly, our proposed approach, \method, consistently surpasses all the baseline methods.
We design the behavior of attribute-based item critiquing for \method to provide detailed feedback in the interaction process, which can help CRSs generate diverse and relevant item recommendations.
In addition, we design the behavior of generative item scoring, which can effectively leverage the text generation capabilities of LLMs to align closely with the real user preference to give accurate item scores.
Notably, \method achieves a substantial performance improvement on the low-resource \textsc{INSPIRED} dataset.
This is because we focus on optimizing the simulated user, which provides feedback on recommended items instead of directly generating recommendations.
This design allows us to utilize other datasets (\eg \textsc{ReDial}) for training, effectively mitigating the challenge of data scarcity.
For this point, we provide a detailed analysis in Section~\ref{sec:transferability}.
Furthermore, our approach employs a general LLM as the CRS without fine-tuning it on the datasets. The results demonstrate significant performance gains compared to the original LLM.
(\ie \method vs. \texttt{Phi-4}; more results in Section~\ref{sec:llm}).
This highlights that general LLMs inherently possess substantial knowledge about target items, and the interaction with our simulated user effectively aligns them with target application distributions, thereby unlocking their potential as high-performing CRSs.

\subsection{Detailed Analysis}

\subsubsection{Ablation Study}

\begin{figure}[t]
    \centering
    \includegraphics[width=0.49\linewidth]{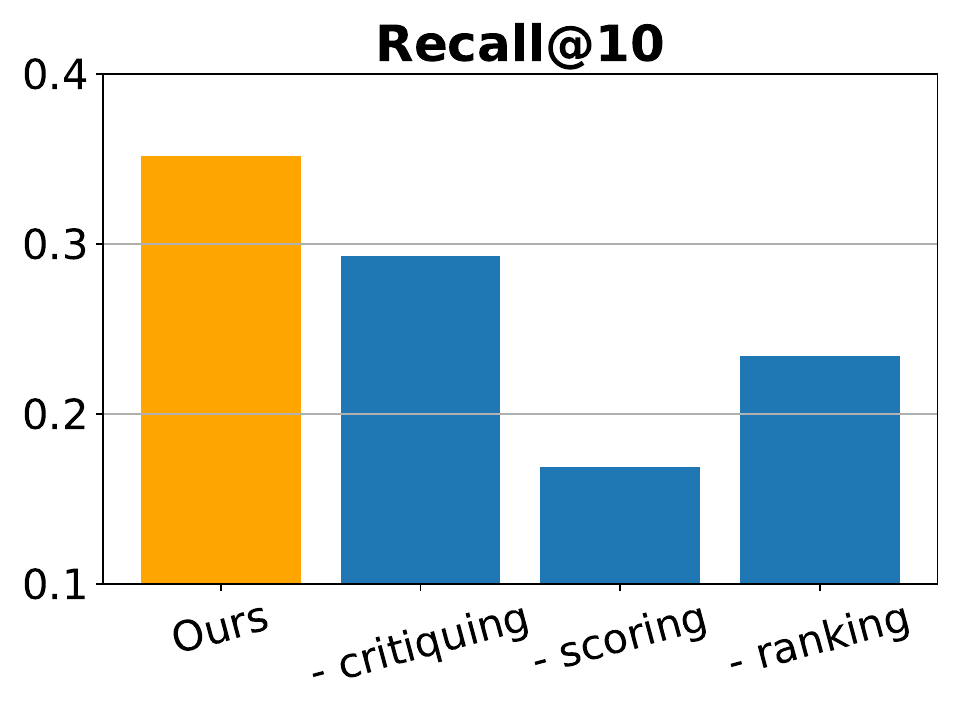}
    \includegraphics[width=0.49\linewidth]{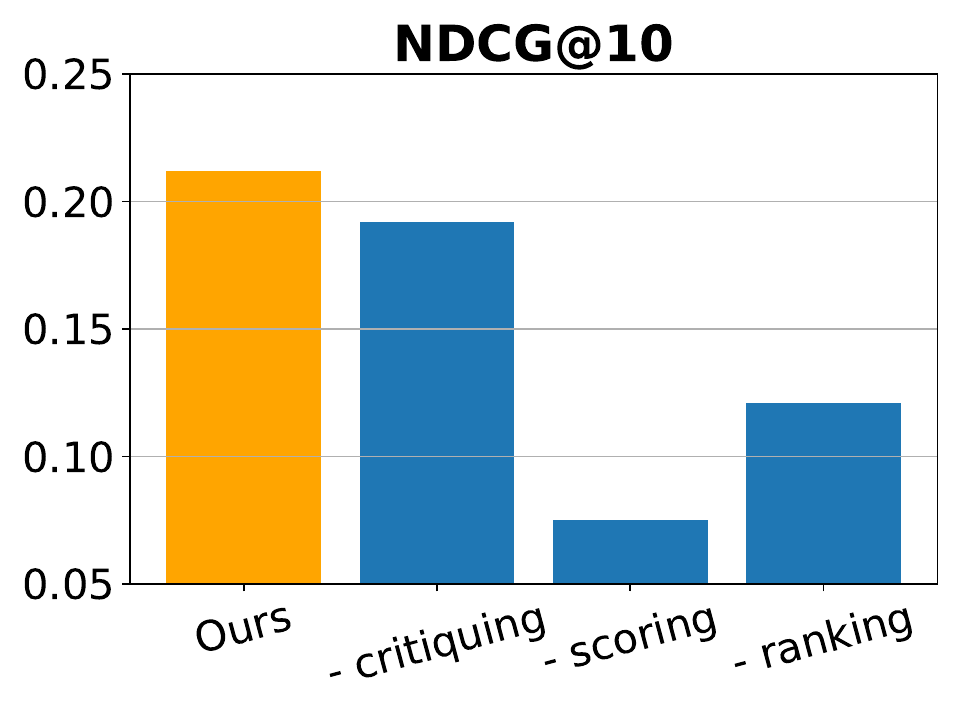}
    \caption{
        Ablation study on the \textsc{INSPIRED} dataset.
        ``critiquing'' refers to attribute-based item critiquing, ``scoring'' refers to generative item scoring, and ``ranking'' refers to efficient candidate ranking.
    }
    \label{fig:ablation}
\end{figure}

Our approach incorporates several components to enhance the interaction between the CRS and the simulated user, ultimately improving recommendation performance.
To verify the effectiveness of each component, we conduct an ablation study on the \textsc{INSPIRED} dataset using \texttt{Phi-4} as the CRS.
The evaluation metrics include Recall@10 and NDCG@10, as the length of all item lists is fixed at 10.
The ablation study considers the removal of specific user behaviors, namely attribute-based item critiquing and generative item scoring. 
Additionally, we examine the impact of excluding the efficient ranking method.

The results are shown in Figure~\ref{fig:ablation}.
We can see that removing any component would lead to a drop in the performance.
It indicates that all the components in our approach are useful to improve the performance of recommendation.
Notably, the removal of generative item scoring results in the most significant performance drop. This finding underscores the importance of generative item scoring in our approach, as it predicts user acceptance of recommended items and directly influences CRS performance.
For the two metrics Recall@10 and NDCG@10, the impact is more significant for NDCG@10.
This is attributed to the focus of NDCG on the ranking quality of ground-truth items, which depends heavily on the generative item scoring behavior.

\subsubsection{Implementation of User Behaviors}

\begin{table}[t]
    \centering
    \caption{
    Ablation study of the two behaviors on the \textsc{INSPIRED} dataset. 
    The best one is marked in bold. 
    Numbers marked with * indicate that the improvement is statistically significant compared with the baseline (t-test
    with p-value < 0.05). 
    }
    \resizebox{\columnwidth}{!}{%
    \begin{tabular}{l|cccc}
        \toprule \textbf{Methods}                                                            & \textbf{Recall@10}       & \textbf{Recall@50}       & \textbf{NDCG@10}         & \textbf{NDCG@50}         \\
        \midrule \method                                                                & \textbf{0.352}* & \textbf{0.559}* & \textbf{0.212}* & \textbf{0.258}* \\
        \midrule \midrule \multicolumn{5}{c}{\textit{Attribute-based item critiquing}} \\
        \midrule \midrule - instruction tuning                                      & 0.300           & 0.511           & 0.184           & 0.232           \\
        % - critique                                                                  & 0.331           & 0.525           & 0.193           & 0.236           \\
        - attribute                                                                 & 0.299           & 0.487           & 0.169           & 0.212           \\
        \midrule \midrule \multicolumn{5}{c}{\textit{Generative item scoring}}       \\
        \midrule \midrule - instruction tuning                                      & 0.156           & 0.382           & 0.083           & 0.131           \\
        discriminative scoring & 0.287 & 0.503 & 0.124 & 0.191 \\
        majority voting                                                               & 0.189           & 0.387           & 0.093           & 0.136           \\
        {\begin{tabular}[c]{@{}l@{}}majority voting +\\generative scoring\end{tabular}}        & 0.205           & 0.486           & 0.095           & 0.158           \\
        \bottomrule
    \end{tabular}%
    } \label{tab:user-model}
\end{table}

For our simulated user \method, we design two user behaviors for it.
In the above part, we have verified the effectiveness of both behaviors.
In this part, we further justify our implementations by comparing them with alternative designs.
Specifically, we conduct the study on the \textsc{INSPIRED} dataset using \texttt{Phi-4} as the CRS.
The evaluation metrics include Recall@10, Recall@50, NDCG@10, and NDCG@50.
For attribute-based item critiquing, we consider alternative implementations, including removing instruction tuning and item attributes.
For generative item scoring, we consider variations including removing instruction tuning, adopting majority voting, discriminative scoring, and combining majority voting with generative item scoring (\ie using the scores from generative item scoring as weights for majority voting).

The results are shown in Table~\ref{tab:user-model}.
We can see that any alternative implementation would lead to performance degradation.
It indicates the effectiveness of our implementations for the two user behaviors.
For attribute-based item critiquing, the most significant performance decline occurs when item attributes are removed. 
This highlights the importance of attributes in item critiques, as they encapsulate specific user preference information that CRSs can leverage to recommend more relevant items.
For generative item scoring, the largest performance drop is observed when instruction tuning is removed. 
This finding emphasizes the critical role of alignment with real user preferences in item scoring, consistent with observations from a recent work~\cite{he2024reindex}.
Furthermore, generative item scoring significantly outperforms majority voting and discriminative scoring, demonstrating the distributional gap between LLMs and real users.
Thus, even if we incorporate the scores given by generative item scoring into majority voting, there still exists a significant performance gap with our implementation.

\subsubsection{Transferability of the Simulated User}
\label{sec:transferability}

\begin{table}[t]
    \centering
    \caption{
    Transferability of the simulated user on the \textsc{INSPIRED} dataset. 
    The simulated user is trained using three configurations: the high-resource \textsc{ReDial} dataset, the low-resource \textsc{INSPIRED} dataset, and a combined dataset incorporating both. 
    The best one is marked in bold. 
    Numbers marked with * indicate that the improvement is statistically significant compared with the baseline (t-test with p-value < 0.05). 
    }
    \resizebox{\columnwidth}{!}{%
    \begin{tabular}{l|cccc}
        \toprule 
        {\begin{tabular}[c]{@{}l@{}}\textbf{Training}\\\textbf{datasets}\end{tabular}}                                                   & \textbf{Recall@10}       & \textbf{Recall@50}       & \textbf{NDCG@10}         & \textbf{NDCG@50}         \\
        \midrule 
        {\begin{tabular}[c]{@{}l@{}}INSPIRED\\(low resource)\end{tabular}}   & 0.234           & 0.463           & 0.127           & 0.178           \\
        {\begin{tabular}[c]{@{}l@{}}ReDial\\(high resource)\end{tabular}} & 0.304           & 0.539           & 0.176           & 0.229           \\
        \midrule ReDial+INSPIRED                                                     & \textbf{0.352}* & \textbf{0.559}* & \textbf{0.212}* & \textbf{0.258}* \\
        \bottomrule
    \end{tabular}%
    } \label{tab:transfer}
\end{table}

One important limitation of existing CRSs is their requirement for specialized training tailored to each dataset due to the differences in item sets.
Our simulated user provides a potential solution to this issue.
This is because it only needs to provide feedback on recommended items instead of directly generating item recommendations.
To validate the transferability of our simulated user from high-resource datasets to low-resource datasets, we conduct experiments on the \textsc{ReDial} and \textsc{INSPIRED} datasets using \texttt{Phi-4} as the CRS.
The evaluation metrics include Recall@10, Recall@50, NDCG@10, and NDCG@50.
Specifically, we train the simulated user under three configurations: using only the high-resource \textsc{ReDial} dataset, using only the low-resource \textsc{INSPIRED} dataset, and using a combined dataset that integrates both.
The performance is then assessed on the test set of the low-resource \textsc{INSPIRED} dataset.

The results are shown in Table~\ref{tab:transfer}.
As we can see, the simulated user trained only on the high-resource \textsc{ReDial} dataset performs much better than directly using the low-resource \textsc{INSPIRED} dataset itself.
This substantial improvement underscores the transferability of our simulated user across different datasets.
Furthermore, incorporating the \textsc{INSPIRED} dataset can further improve the performance.
This indicates that the simulated user can effectively utilize the data from diverse sources in training.
We leave the scaling of training data for CRSs and simulated users as future work.

\subsubsection{Different LLMs as CRSs}
\label{sec:llm}

\begin{table}[t]
    \centering
    \caption{
        The performance of various LLMs as the CRS with and without our simulated user \method on the \textsc{INSPIRED} dataset.
        The best method using the same model is marked in bold.
        Numbers marked with * indicate that the improvement is statistically significant compared with the baseline (t-test with p-value < 0.05).
    }
    \resizebox{\columnwidth}{!}{%
    \begin{tabular}{c|c|cccc}
        \toprule
        \textbf{Models}                                   & \textbf{Methods}    & \textbf{Recall@10} & \textbf{Recall@50} & \textbf{NDCG@10} & \textbf{NDCG@50} \\
        \midrule
        \multirow{2}{*}{\begin{tabular}[c]{@{}c@{}}Llama-3.1-\\8B-Instruct\end{tabular}}  & zero-shot & 0.166     & 0.285     & 0.090   & 0.117   \\
                                                & +\method      & \textbf{0.324}* & \textbf{0.529}* & \textbf{0.194}* & \textbf{0.240}* \\
        \midrule
        \multirow{2}{*}{Phi-4}                  & zero-shot & 0.145     & 0.277     & 0.077   & 0.106   \\
                                                & +\method      & \textbf{0.352}* & \textbf{0.559}* & \textbf{0.212}* & \textbf{0.258}* \\
        \midrule
        \multirow{2}{*}{\begin{tabular}[c]{@{}c@{}}Llama-3.3-\\70B-Instruct\end{tabular}} & zero-shot & 0.166     & 0.311     & 0.093   & 0.127   \\
                                                & +\method      & \textbf{0.338}* & \textbf{0.551}* & \textbf{0.219}* & \textbf{0.268}* \\
        \midrule
        \multirow{2}{*}{GPT-4o}                 & zero-shot & 0.217     & 0.378     & 0.124   & 0.159   \\
                                                & +\method      & \textbf{0.317}* & \textbf{0.549}* & \textbf{0.196}* & \textbf{0.249}*   \\
        \bottomrule
    \end{tabular}%
    }
    \label{tab:llm}
\end{table}

In our main experiments, we use \texttt{Phi-4} as the LLM-based CRS due to its strong performance and suitable model size.
In this part, we investigate whether our approach is applicable to other LLMs with varying capabilities.
Specifically, we evaluate a diverse set of open-source and closed-source LLMs, including \texttt{Llama-3.1-8B\-Instruct}, \texttt{Llama-3.3-70B-Instruct}, and \texttt{GPT-4o}.
We report the results of Recall@10, Recall@50, NDCG@10, and NDCG@50 on the \textsc{INSPIRED} dataset.

As shown in Table~\ref{tab:llm}, our approach can effectively improve the performance of all the evaluated LLM-based CRSs by a large margin.
It demonstrates the general applicability of our method to various LLM-based CRSs.
The improvement can be attributed to our strategy for synthesizing instruction data, which incorporates two types of item lists (\ie, lists with and without ground-truth items).
With this strategy, our trained simulated user can effectively handle recommended item lists generated by different LLMs.

\subsubsection{Impact of the Search Strategy}

\begin{table}[t]
    \centering
    \caption{
        The performance of various search methods with the simulated user on the \textsc{INSPIRED} dataset.
        The best one is marked in bold.
    }
    \resizebox{\columnwidth}{!}{%
    \begin{tabular}{l|cccc}
        \toprule 
        \textbf{Search methods}          & \textbf{Recall@10}      & \textbf{Recall@50}      & \textbf{NDCG@10}        & \textbf{NDCG@50}        \\
        \midrule 
        No search (zero shot)   & 0.145          & 0.277          & 0.077          & 0.106          \\
        \midrule
        Monte Carlo search      & 0.293          & 0.484          & 0.192          & 0.237          \\
        Greedy search (small)   & 0.317          & 0.436          & 0.196          & 0.223          \\
        Greedy search (large)   & 0.338          & \textbf{0.580} & 0.210          & \textbf{0.265} \\
        Beam search (ours)      & \textbf{0.352} & 0.559          & \textbf{0.212} & 0.258          \\
        \bottomrule
    \end{tabular}%
    } \label{tab:search}
\end{table}

Recall that in Section~\ref{sec:interaction}, we employ beam search to facilitate the interaction between CRSs and our simulated user. To systematically investigate the impact of search strategies within our approach, we compare beam search with several alternative search strategies, including Monte Carlo search and greedy search.
Monte Carlo search performs multiple one-step sampling operations.
That is, it limits the search depth to one.
Greedy search chooses the best among multiple samples at each step.
That is, it limits the beam width to one.
For all the search strategies, we use the same formulation described in Section~\ref{sec:interaction} and only replace beam search with each search method.
To ensure comparability, we maintain the same number of overall generated item lists across all search strategies, thus preserving a similar search space.
Additionally, for greedy search, we examine a variant with the same expand width as beam search, referred to as \textit{greedy search (small)}, alongside the standard version where the number of generated item lists matches beam search, referred to as \textit{greedy search (large)}.
We conduct this comparison on the \textsc{INSPIRED} dataset, utilizing \texttt{Phi-4} as the CRS. 
The evaluation metrics include Recall@10, Recall@50, NDCG@10, and NDCG@50.

The results are shown in Table~\ref{tab:search}.
As we can see, all the search methods significantly outperform the no-search baseline, suggesting the importance of multi-turn interaction for CRSs.
When maintaining a similar search space (\ie Monte Carlo, greedy (large), and beam search), we observe comparable performance for greedy and beam search.
It indicates the effectiveness of our efficient ranking method, which takes advantage of all the searched items at a low cost.
In addition, they perform better than Monte Carlo search.
This is because the search depth of Monte Carlo search is limited to one, which does not involve the feedback of attribute-based item critiquing from our simulated user.
Furthermore, when we set the expand width of greedy search to the same as beam search, we find a drop in its performance.
It demonstrates the importance of search space for effective interaction.

\subsubsection{Efficiency of Search-Based Interaction}

\begin{figure}[t]
    \centering
    \includegraphics[width=0.49\linewidth]{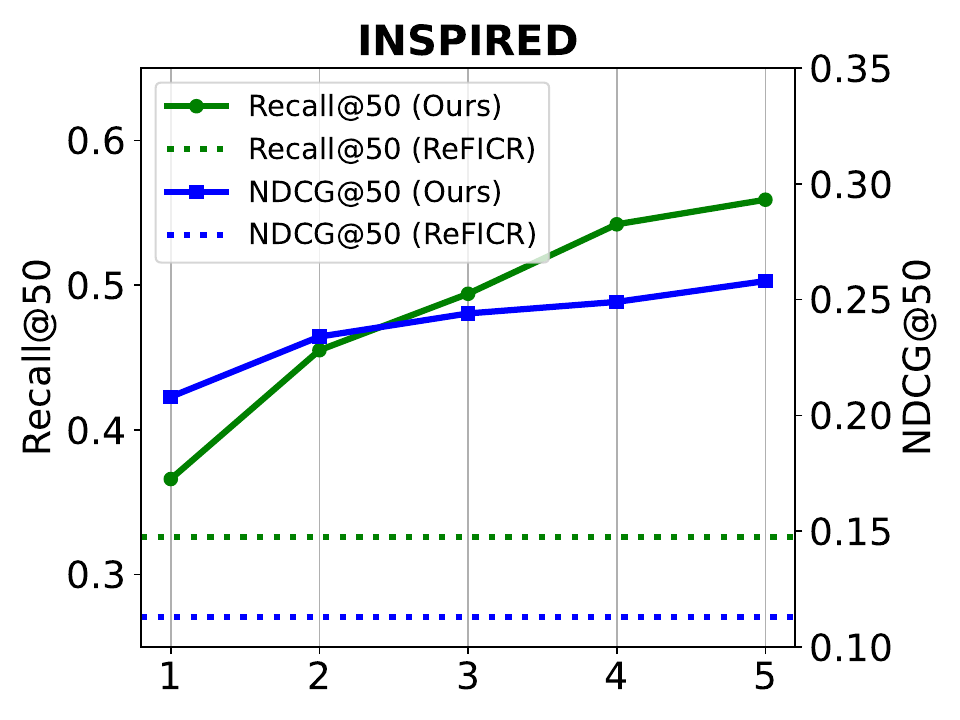}
    \includegraphics[width=0.49\linewidth]{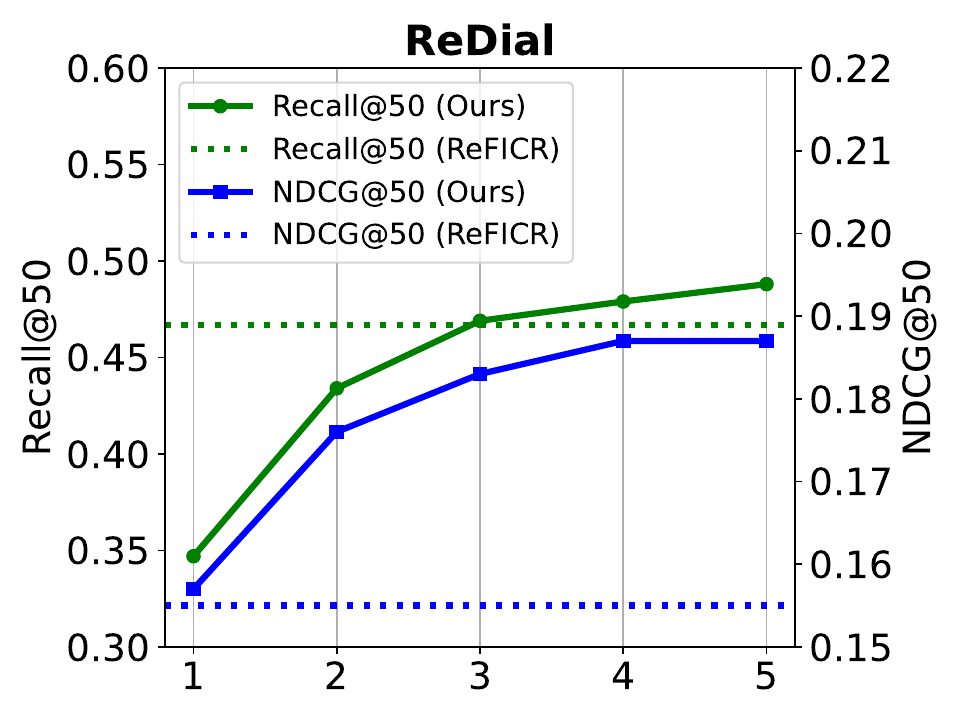}
    \caption{
        The performance changes with respect to the depth of search.
        As a reference, we add the performance of ReFICR.
    }
    \label{fig:search}
\end{figure}

In addition to the superior performance, the efficiency is also vital to make the computation cost of the search manageable.
Thus, we analyze the search efficiency by plotting performance changes with respect to the search depth.
Specifically, we conduct the study on the \textsc{ReDial} and \textsc{INSPIRED} datasets using \texttt{Phi-4} as the CRS.
We report Recall@50 and NDCG@50 metrics across search depths ranging from one to five.
For comparison, we include the performance of the best baseline method, ReFICR, as a reference point.

As illustrated in Figure~\ref{fig:search}, we observe a continuous performance improvement on both datasets.
It indicates the effectiveness of the attribute-based item critiquing behavior of our simulated user, which can consistently provide valuable feedback for CRSs to revise recommendations.
Compared with ReFICR, our approach achieves better performance at the first and third turns on the \textsc{INSPIRED} and \textsc{ReDial} datasets, respectively.
Even within a limited cost, our method achieves superior performance, underscoring its efficiency.

\section{Conclusion}

In this paper, we propose a generative reward model based simulated user, named \textbf{\method}, for search-based interaction with CRSs to solve the issue of complex user preferences.
First, we take inspiration from generative reward models to design two actions for our simulated user: \ie \textit{generative item scoring} as coarse-grained feedback and \textit{attribute-based item critiquing} as fine-grained feedback.
Then, we unify these two actions into the format of instruction, which enables us to develop a unified simulated user with instruction tuning on synthesized data.
Based on this simulated user, automatic interaction is conducted with CRSs to help them understand complex user preferences and revise recommendations with feedback of different granularity.
Furthermore, we leverage beam search for the interaction process to achieve a balance between effectiveness and efficiency.
Finally, we propose an \textit{efficient} candidate ranking method to improve the recommendation results derived from the interaction process.
Extensive experiments demonstrate the effectiveness, efficiency, and transferability of our approach.

As future work, we will investigate the scaling of model size and dataset size for simulated users in conversation recommendation.

\begin{acks}
This work was partially supported by National Natural Science Foundation of China under Grant No. 92470205 and 62222215, Beijing Municipal Science and Technology Project under Grant No. Z231100010323009, and Beijing Natural Science Foundation under Grant No. L233008, and the Outstanding Innovative Talents Cultivation Funded Programs 2022 of Renmin University of China.
Xin Zhao is the corresponding author.
\end{acks}

% \clearpage
\bibliographystyle{ACM-Reference-Format}
\balance
\bibliography{ref.bib}

\appendix

\end{document}